# THERMAL BENCHMARK AND POWER BENCHMARK SOFTWARE

*Marius Marcu, Mircea Vladutiu, Horatiu Moldovan and Mircea Popa*

Department of Computer Science, “Politehnica” University of Timisoara, Faculty of Automation and Computer Science and Engineering, 2 V. Parvan, Timisoara, Romania, Phone: (40) 256-403263, Fax: (40) 256-403214, E-Mail: marius.marcu@cs.upt.ro, WWW: http://www.cs.upt.ro/~mmarcu

**ABSTRACT**

Power consumption and heat dissipation become key elements in the field of high-end integrated circuits, especially those used in mobile and high-speed applications, due to their increase of transistor count and clock frequencies. Dynamic thermal management strategies have been proposed and implemented in order to mitigate heat dissipation. However, there is a lack of a tool that can be used to evaluate DTM strategies and thermal response of real life systems. Therefore, in this paper we introduce and define the concepts of thermal benchmark software and power benchmark software as a software application for run-time system level thermal and power characterization.

## 1. INTRODUCTION

High-end integrated circuits’ power consumption is continuing increasing, as a consequence of the integration area increase, number of integrated units increase and working frequencies increase. Because of the increase in power consumption, new problems arise, which are related to the limitation of the temperature at the semiconductors’ level, in order to avoid the physical damages of the circuits [1, 2]; and battery saving for mobile systems. This is the reason why the temperature and power consumption is addressed often during the different phases of developing the component parts of the computing systems [2].

Decreased operating speed and decreased reliability are the most important consequences of high operating temperatures [2, 3] that may cause software errors or physical damage of the circuit. Therefore, high level integrated circuits are designed to operate reliably within a defined temperature range. Outside of this range, there is no assurance that the integrated circuits will continue to function correctly [2]. Hence, in the context of increasing in temperature, thermal management methods have to be applied. The ultimate goal of these techniques is to keep the circuit at or below its maximum operating temperature regardless of running application or environment temperature.

Proper thermal management depends on two major elements: thermal packaging and thermal management. Thermal packaging includes heatsink properly mounted to the processor and effective fans directing airflow through the system chassis [2,3,4]. Dynamic thermal management (DTM) strategies are used to reduce packaging cost without unnecessarily limiting performance, the package is designed for the worst typical application and any application that dissipate more heat should activate an alternative, run-time thermal management technique [3,4,5]. This method is already implemented in modern microprocessors [4].

As a conclusion for thermal management, DTM schemes are designed as solutions to deal with the worst-case applications while the thermal package deals with the average or typical applications. Usually the proposed DTM strategies are mainly evaluated by simulation or using performance benchmarks (SPEC CPU2000 [7]). The work presented in this article propose two new benchmark applications, we called **thermal benchmark** and **power benchmark**. Thermal benchmark concept is not new but is not used in the context of thermal management. The context found for thermal benchmark is the work of Szekely et. all [6], regarding a thermal benchmark circuit, designed for thermal characterization of circuit level management. But there is no reference, as we know at this moment, about thermal benchmark software, therefore we intend to define and validate such one here. We intend also to introduce and define here the power benchmark software for battery powered systems. Power benchmark concept it is not new but it is recently proposed by SPEC [8]. For now, the SPEC Power-Performance Committee is confining its research only into small-sized and medium-sized servers. While the new "SPECpower" benchmark has not been defined, SPEC expects the benchmark to be in place by the first quarter of 2007 [8,9]. Therefore we can say that we propose a new power benchmark for battery powered devices, that may be called also battery benchmark.

Thermal benchmark applications are benchmarks proposed to evaluate and/or calibrate DTM solutions,

thermal hardware sizing related to applications the system is expected to support, on-line thermal testing and monitoring, system level thermal control. Power benchmark applications are a new kind of benchmarks proposed to evaluate and/or calibrate power management. The new power or battery benchmark also indicates the importance of power consumption in evaluating new compute platforms and battery devices.

The paper is organized as follows. In the next section we define the thermal benchmark concept and its implementation. Section 3 of this paper describes our power benchmark software and its implementation. Experimental results and conclusions are presented in Section 5 and Section 6 respectively.

## 2. THERMAL BENCHMARK DEFINITION

A computer benchmark is typically a computer program that performs a strictly defined set of operations (a workload) and returns some form of result (a metric) describing how the tested computer performed. Computer benchmark metrics usually measure speed (how fast was the workload completed) or throughput (how many workloads per unit time were measured). Running the same computer benchmark on multiple computers allows a comparison to be made [7].

We propose to extend the concept of benchmarking with a new metric: the temperature, and we name it thermal benchmark.

**Definition 1** - Thermal benchmark is defined as a software program that describes the thermal behavior of the system or component with respect to certain stimulus (workload).

A thermal benchmark must by able to distinguish the way a hardware device temperature is increasing with workload and the way it's temperature decrease when the workload is finished. Therefore, a thermal benchmark is composed by three components (Fig. 1):

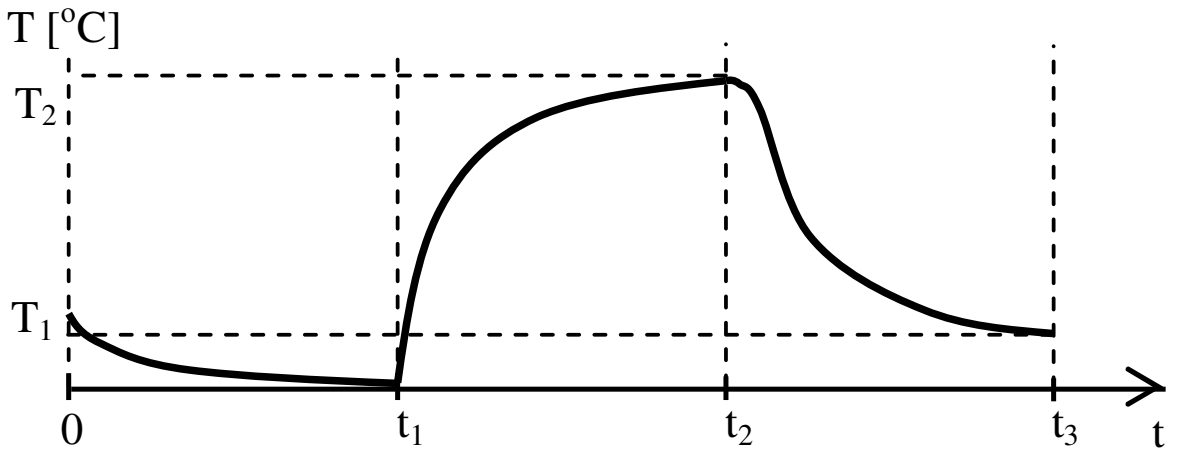

Fig. 1 Thermal benchmark definition

- the first range [0-$t_1$), is intended for idle mode temperature. In this step, the component does not execute anything, but the power saving mechanisms are prevented to occur.
- the second range [$t_1$-$t_2$) represents the warming phase, when a certain workload is executed. SPEC CPU2000 or any type of other applications can be executed as workload.
- the last range [$t_2$-$t_3$) represents the cooling phase intended for the component to reach again the idle state temperature. In this step, the component does not execute anything, but the power saving mechanisms are prevented to occur.

There are two ways the thermal benchmark can be implemented:

- fixed times thermal benchmark – the benchmark times $t_1$, $t_2$ and $t_3$ are predefined and constant. This kind of thermal benchmark shows the maximum component temperature when a certain workload is applied for a constant period of time ($t_2$-$t_1$);
- fixed temperatures thermal benchmark – the benchmark times are variable and benchmark temperatures are predefined and constant. This benchmark shows the time interval needed to the component to reach a certain temperature when a certain type of workload is applied. Usually $T_2$ can be selected to be the maximum temperature ($T_{max}$) reached by component and $T_1$ can be the idle state temperature ($T_{idle}$).

We adopted a graphical representation for thermal benchmark, but numerical metrics can be also used. We propose two such metrics: temperature difference – defined as $(T_2-T_1)/(t_2-t_1)$ and temperature area – defined as the area under temperature signature in time interval [$t_1$,$t_2$) (warming temperature area) or [$t_2$,$t_3$) (cooling temperature area). In the following presentation will continue to used the graphical representation because is more significant.

A thermal benchmark can be applied to every hardware device with built-in thermal monitoring capability (such as microprocessor, hard disk or video). Microprocessor thermal benchmark was implemented in our work and this will be further presented.

**Definition 2** – Microprocessor thermal benchmark defined as a software program that describes the thermal behavior of the microprocessor when executing certain workload.

The majority of microprocessors produced in last years include at least one built-in thermal sensor to aid in thermal management of servers, workstation or portable systems. This thermal sensor is connected to a thermal diode on the processor core, therefore it provides the earliest indication of thermal variation that can be read through ACPI or SMBus [4]. More detailed information about microprocessor thermal benchmark implementation and temperature sensors reading from applications running under Windows NT based operating systems was published by the same author in [10,11].

## 3. POWER BENCHMARK DEFINITION

We propose also to extend the concept of benchmarking with another metric: the power consumption, and we name it power benchmark or battery benchmark.

**Definition 3** - Power benchmark is defined as a software program that describes the power consumption of the system with respect to certain stimulus (workload).

A power benchmark must by able to distinguish the way power consumption is increasing with workload related to idle state consumption. Therefore, we define a power benchmark to be composed by three intervals (Fig. 2):

- idle state consumption $[0,t_1)$;
- workload consumption $[t_1,t_2)$;
- post-workload idle consumption $[t_2,t_3)$.

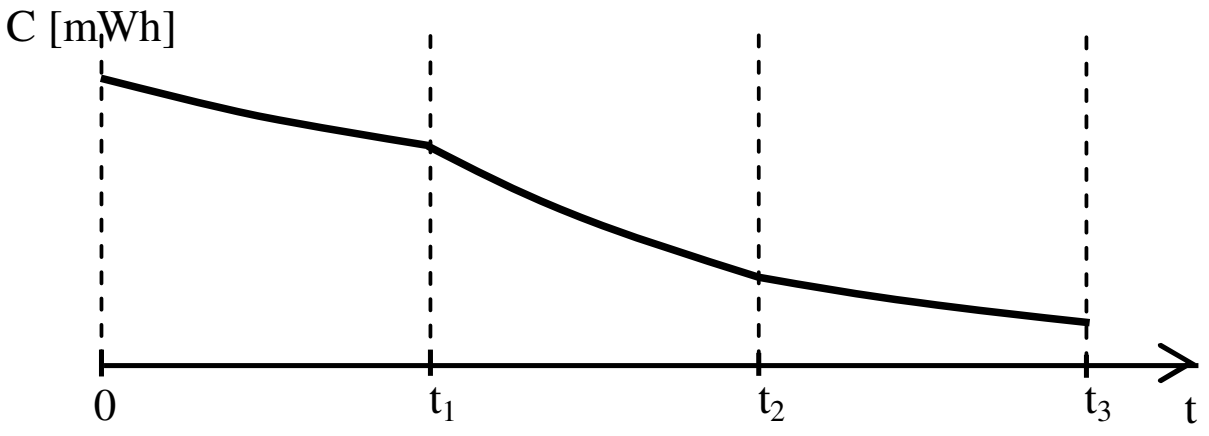

Fig. 2 Power benchmark definition

Some power consumption measures can be used in benchmark: current battery capacity [mWh], discharge rate [W], time remaining [s] or battery capacity percentage [%]. Metrics used for power benchmark are: number of workload operations per battery capacity difference and the discharge time between two battery capacity values. Considering $t_1 = 0$ and $t_3=\infty$, battery discharge profile can be also obtained.

## 4. BENCHMARKS EXPERIMENTAL RESULTS

In order to validate our proposed benchmark software solutions we elaborate a set of test cases, the most important of them being presented in this section. Tests were run a large number of hours on 6 different hardware systems:

- AMD Athlon 1800 MHz, 512 MB RAM, chipset nVidia-nForce2;
- AMD Athlon 1200 MHz, 256 MB RAM, chipset and thermal monitor VIA686 (3 systems);
- AMD Duron 800 MHz, 256 MB RAM, chipset and thermal monitor VIA686;
- Intel Pentium II 400 MHz, 128 MB RAM, chipset Intel 82801AA/ICH, thermal monitor ADM 1025;
- Intel Pentium IV 2800 MHz, 512 MB RAM, chipset Intel 82801EB/ICH5, thermal monitor ADM 1027;
- Intel Pentium III 1000 MHz laptop with 128 MB RAM;
- Intel Pentium IV mobile 2000MHz laptop with 512 MB RAM, chipset Intel 82801 ICH7.

### 4.1 Thermal benchmark stability

The first test was intended to show that thermal benchmark provides the same results on the same system when a certain workload is applied. Fig. 3 shows two thermal signatures for the same system.

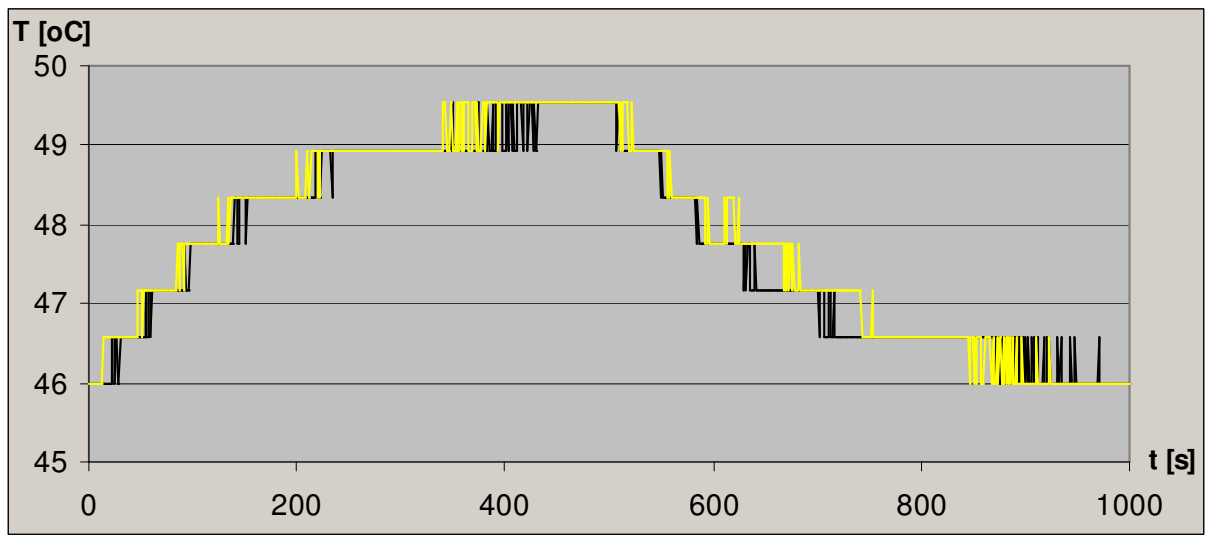

Fig.3 Thermal measurements

Running the benchmark on the same system a large number of times and averaging the measured temperatures we obtains the standard thermal signature of a certain processor for the selected workload (Fig. 4).

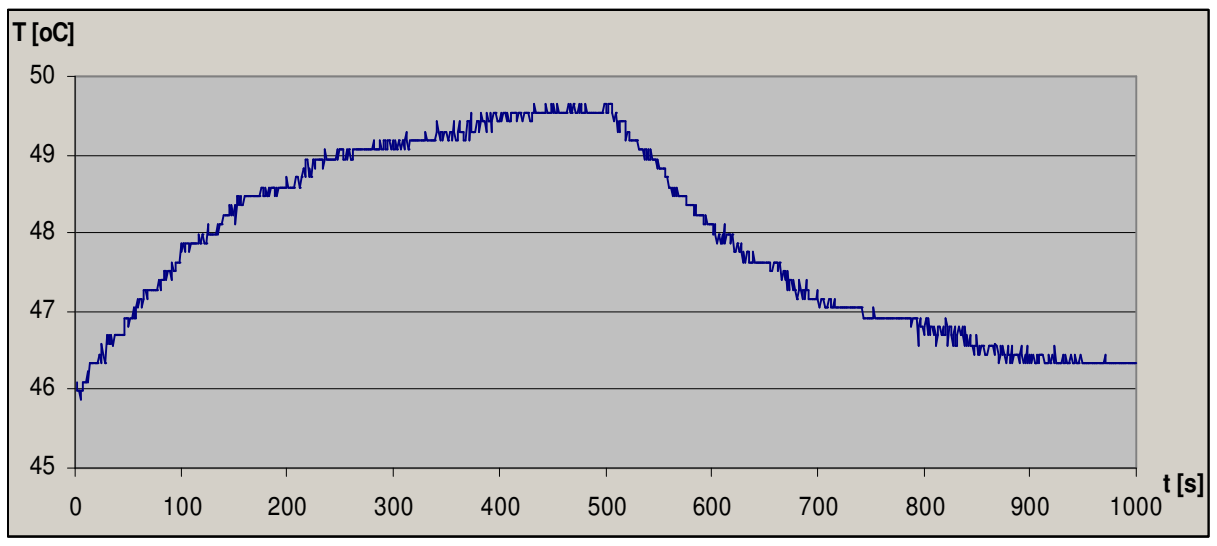

Fig. 4 Thermal signature for a processor

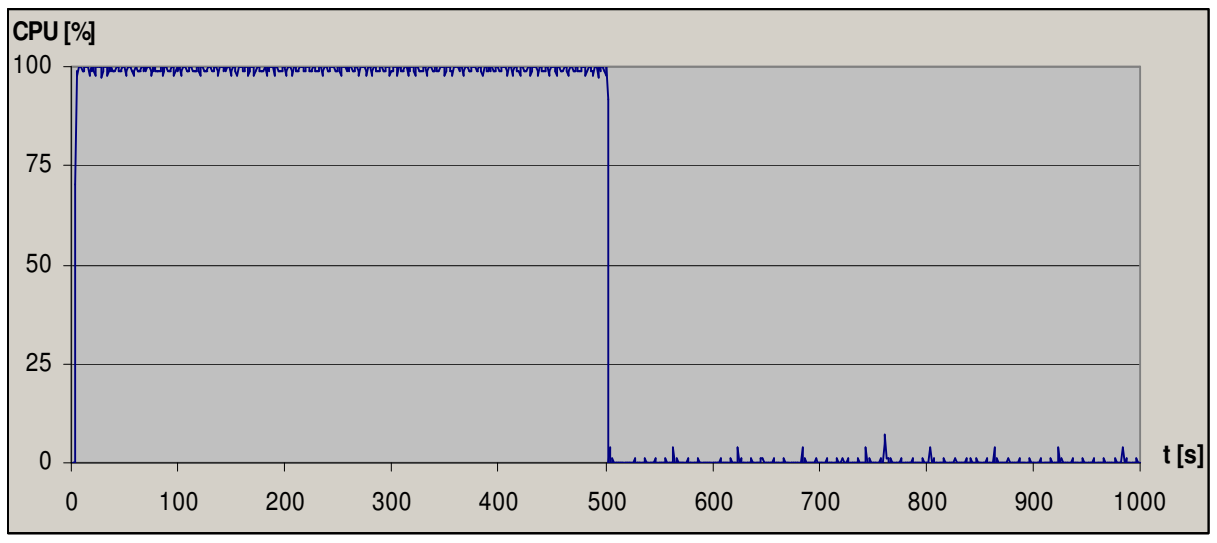

Fig. 5 Processor load when benchmark was applied

### 4.2 Thermal benchmark similarity

The second test was intended to distinguish the thermal signatures for the same type or similar types of processors placed in different systems (case, fans). In Fig. 6 can be observed that thermal response of similar processors is the same for the same workload, but is influenced by cooling components used inside the case. There are different idle state temperatures because of different cooling systems.

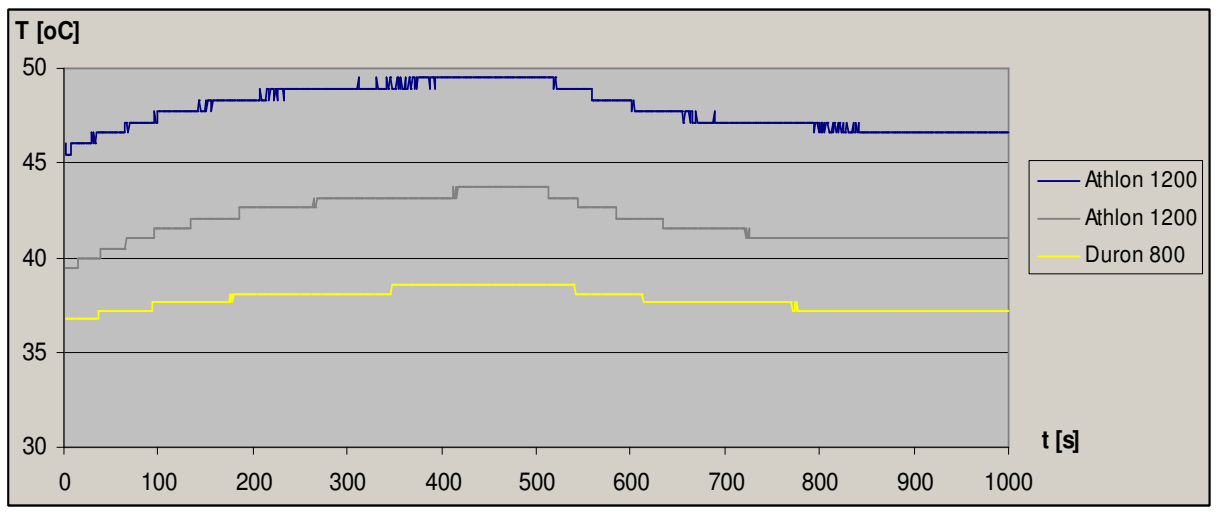

Fig. 6 Thermal benchmark results for similar processors

### 4.3 Thermal benchmark differences

The third test was selected to distinguish between thermal signature of different processors. Fig. 7 shows big differences in thermal responses for Pentium II, Pentium IV and Athlon when the same workload was applied. Intel processors have lower idle state temperatures but when workload is applied temperature increases more dramatically than on AMD processors.

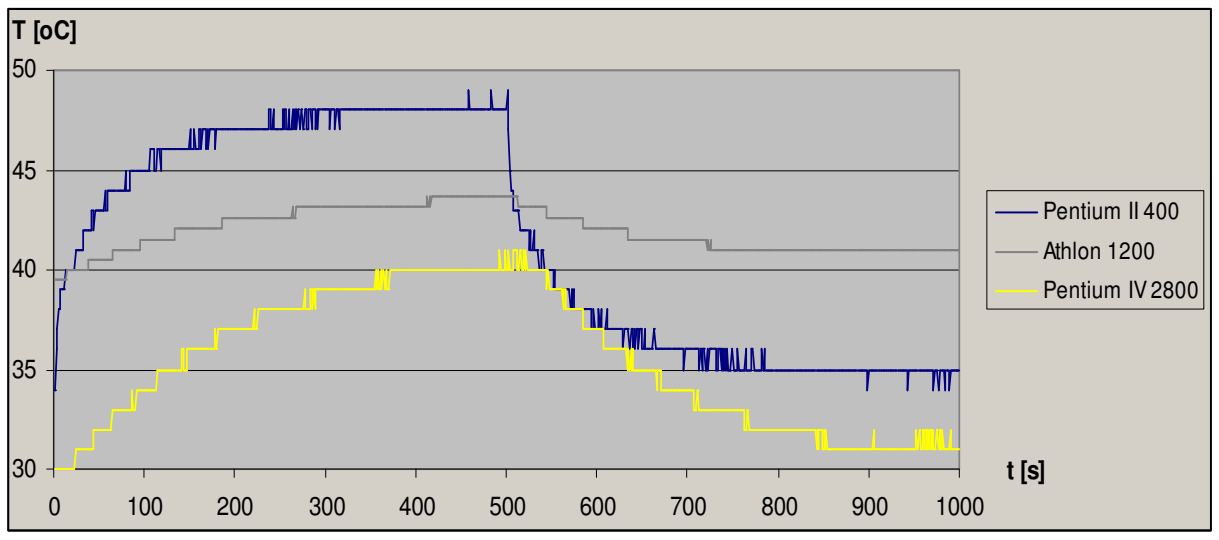

Fig. 7 Thermal benchmark results for different processors

### 4.4 Thermal benchmark workloads

The forth test presents the thermal response of the same processor when different workload types were applied: integer, memory and floating point operations (Fig. 8). Float and memory workloads shows greater increase in temperature than integer workload.

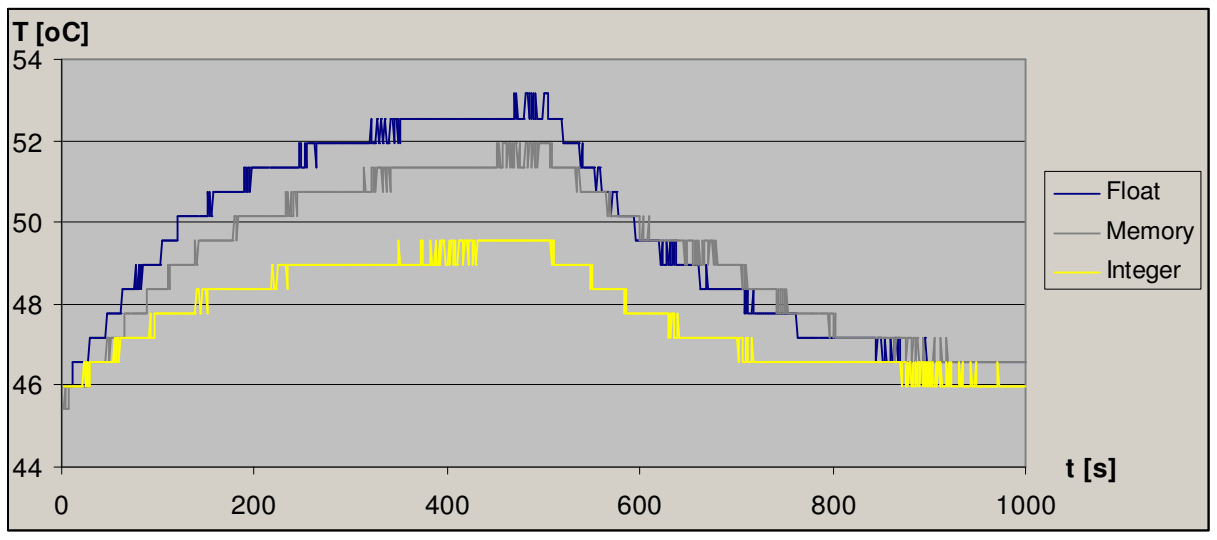

Fig. 8 Thermal response on the same processor for different workloads

### 4.5 Thermal benchmark times

Temperature variation for different thermal benchmark times (100, 200, 300 and 500 seconds) is shown in Fig. 9.

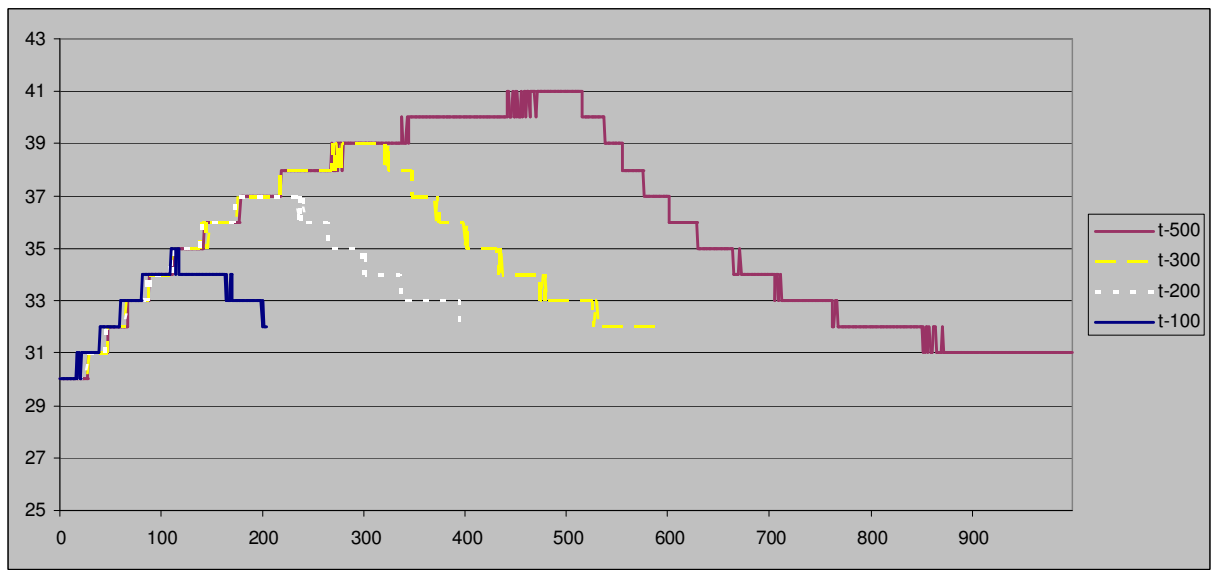

Fig. 9 Thermal response for different times

Using larger periods of benchmark times the maximum temperature for a certain workload is achieved (Fig. 10). The maximum temperature is constant and it can be considered as a propriety of the workload with respect to a certain processor and the attached cooling systems.

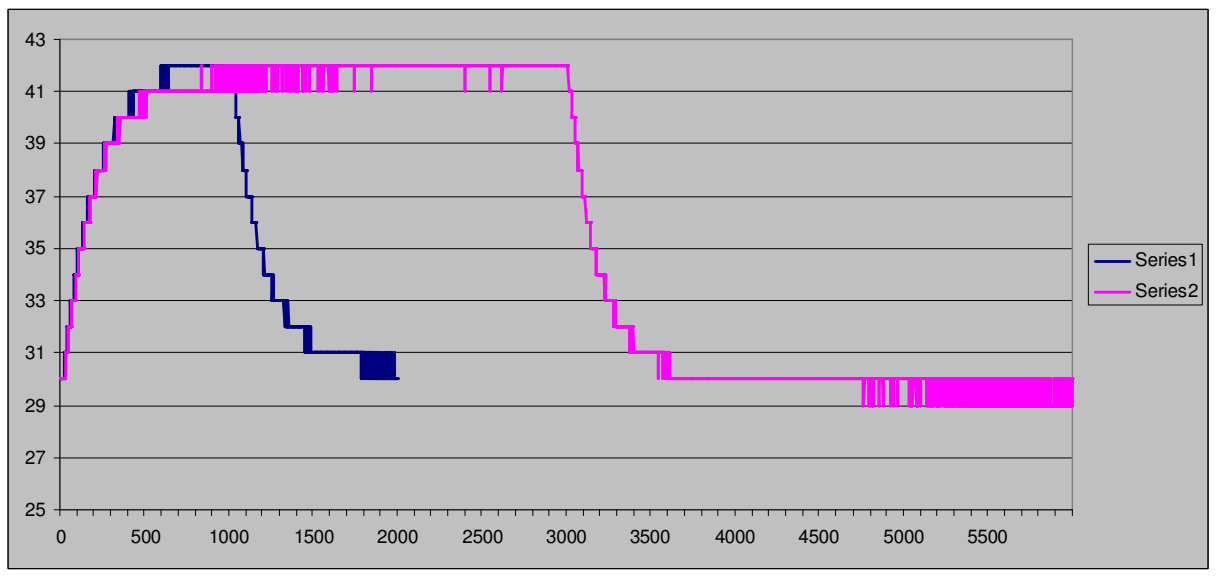

Fig. 10 Maximum temperature for a certain workload

### 4.6 Thermal benchmark threading

Thermal benchmark was run also on two processors with hyper-threading technology, in order to show the influence of number of workload threads on thermal signature. Fig. 11 shows four thermal signatures for one or two workload threads running with and without hyper-threading technology enabled. Two workload threads with hyper-threading enabled have greater impact on temperature increase than other three cases. In case of hyper-threading disabled the maximum temperature is the same and little decrease warm-up time.

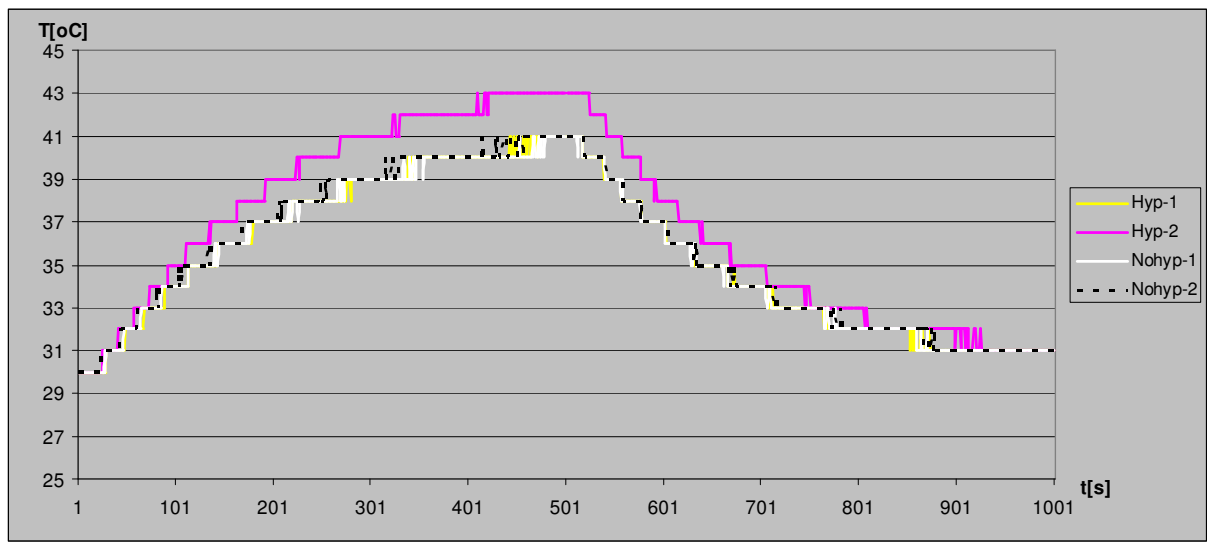

Fig. 11 Thermal signature for different number of workload threads

### 4.7 Microprocessor temperature and motherboard temperature

The influence of workload on the system temperature can be obtained using thermal sensors placed in system case or on the motherboard. Fig. 12 plot the relation between microprocessor thermal signature and internal case thermal signature for a certain workload.

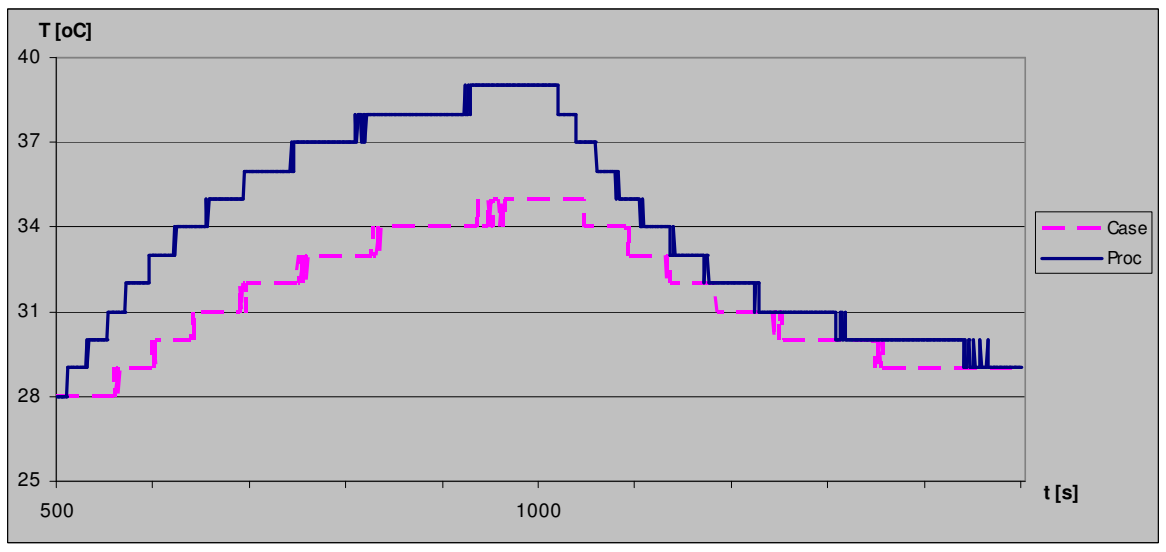

Fig. 12 Microprocessor thermal signature and internal case thermal signature

### 4.8 Power benchmarks

Power benchmark is a software application that monitor battery status when applying certain workload. Battery status is based on several parameters: current battery capacity [mWh], maximum battery capacity [mWh], charge/discharge rate [mW], battery remaining life time [s]. We used these measures to generate power signatures for different system conditions, some of them are further presented (Fig. 13 and 14).

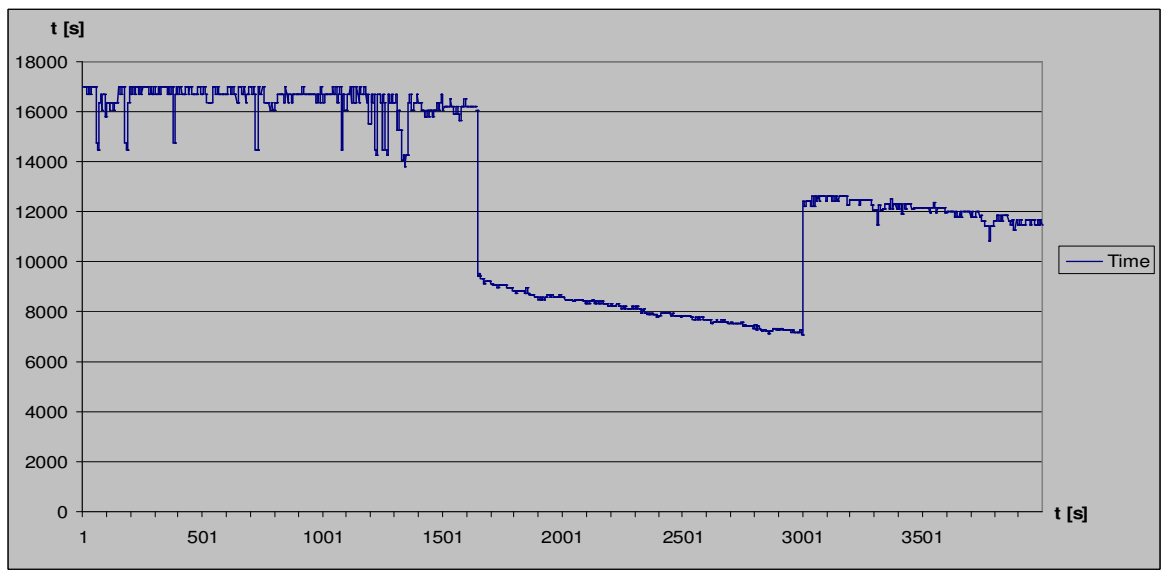

Fig. 13 Remaining lifetime power signature

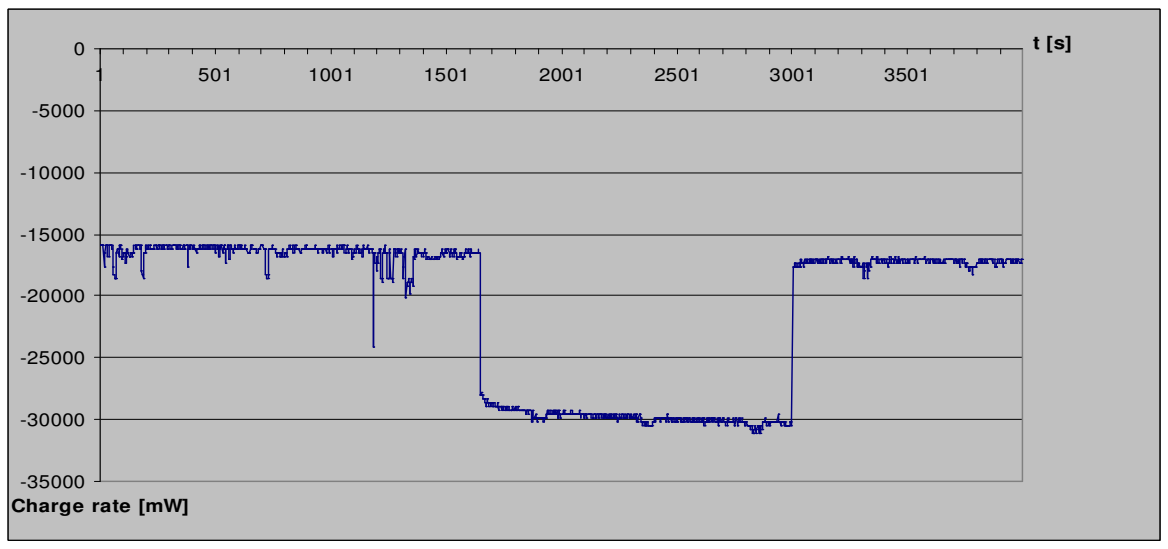

Fig. 14 Discharging rate power signature

Running the power benchmark with different workloads, the draws in Fig. 15 were obtained. In this case there is no significant differences between power signatures of integer, memory or float workloads.

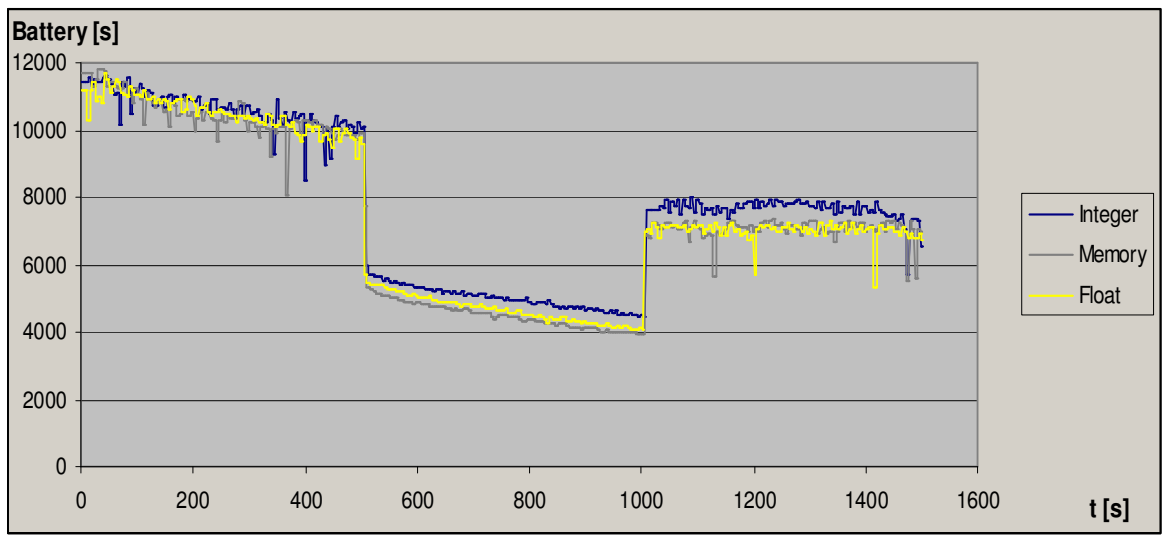

Fig. 15 Power signature for different workloads

## 5. CONCLUSIONS

We have proposed and introduced in this paper two new type of benchmarks: thermal benchmark and power benchmark. Microprocessor thermal benchmarks are software applications proposed to describe thermal behavior of microprocessors when certain types of operations are executed. Thermal benchmarks could have applicability in evaluation and calibration of DTM solutions, thermal hardware sizing related to applications, on-line thermal testing and monitoring, system level thermal control. Microprocessor power benchmarks are software applications that characterize the consumption of the microprocessor related to different types of workloads.

We validated this new concepts by a set of tests presented in this paper.

Future work on thermal and power benchmarking will be directed on selection the right benchmark metrics. Other types of thermal benchmark will be also evaluated: harddisk, case, fans influence. We are also open for further collaborations in order to apply our work to real life testing process or on-line systems testing solutions.